\documentclass[reprint,a4paper,twocolumn,aps,prb,nopacs,superscriptaddress,longbibliography]{revtex4-2}
\usepackage{float}
\usepackage{soul}
\usepackage[english]{babel}
\usepackage[utf8]{inputenc}
\usepackage{fancyhdr}
\usepackage{sidecap}
\usepackage{multirow}

\usepackage{ulem}
\usepackage{graphicx}
\usepackage{braket}
\usepackage{tikz}
\usepackage{color}
\usepackage{amsmath}
\usepackage[colorlinks=True, linkcolor=blue, filecolor=magenta, urlcolor=blue,citecolor=blue]{hyperref}
\usepackage[nameinlink,capitalise]{cleveref}
\usepackage{comment}
\begin{document}

\title{Fermi-liquid behavior and characteristic temperature-dependent susceptibility in clean RuO$_2$ crystal}

\author{Shubhankar Paul}
\affiliation{Toyota Riken–Kyoto University Research Center (TRiKUC), Kyoto 606-8501, Japan}
\affiliation{Department of Electronic Science and Engineering, Graduate School of Engineering, Kyoto University, Kyoto 615-8510, Japan}
\affiliation{Department of Physics, Indian Institute of Technology Kanpur, Kanpur 208016, India}
\author{Atsutoshi Ikeda}
\affiliation{Department of Electronic Science and Engineering, Graduate School of Engineering, Kyoto University, Kyoto 615-8510, Japan}
\author{Hisakazu Matsuki}
\affiliation{Toyota Riken–Kyoto University Research Center (TRiKUC), Kyoto 606-8501, Japan}
\affiliation{Institute for Chemical Research, Kyoto University, Uji, Kyoto 611-0011, Japan}
\author{Giordano Mattoni}
\affiliation{Toyota Riken–Kyoto University Research Center (TRiKUC), Kyoto 606-8501, Japan}
\author {J\"{o}rg Schmalian}
\affiliation{Institute for Theory of Condensed Matter and Institute for Quantum Materials and Technologies, Karlsruhe Institute of Technology, Karlsruhe 76131, Germany}
\author{Kunihiko Yamauchi}
\affiliation{Center for Spintronics Research Network (CSRN), The University of Osaka, Toyonaka, Osaka 560-8531, Japan }
\author{Chanchal Sow}
\affiliation{Department of Physics, Indian Institute of Technology Kanpur, Kanpur 208016, India}
\author{Shingo Yonezawa}
\affiliation{Department of Electronic Science and Engineering, Graduate School of Engineering, Kyoto University, Kyoto 615-8510, Japan}
\author{Yoshiteru Maeno}
\affiliation{Toyota Riken–Kyoto University Research Center (TRiKUC), Kyoto 606-8501, Japan}
\affiliation{Department of Engineering, Kyoto University of Advanced Science (KUAS), Kyoto 615-8577, Japan}
\date{\today}
\begin{abstract}

The magnetic nature of the altermagnet candidate RuO$_2$ remains under debate.
It has been recently shown from quantum oscillations and angle-resolved photoemission spectroscopy (ARPES) that the high-quality RuO$_2$ bulk single crystal is a paramagnetic metal. 
Here we report the specific heat and magnetic susceptibility in ultra-clean RuO$_2$ single crystals with residual resistivity ratio up to 1200. 
The magnetic susceptibility increases with temperature and is phenomenologically fitted with an inclusion of $T\textrm{ln}(T/T_0)$ over a wide temperature range up to 400 K.
In contrast, the energy dependence of the density of states and thermal activation of quasiparticles lead to a decrease with temperature.
Such characteristic temperature dependence, similar to that observed in other $d$-electron metals, is attributable to an enhanced orbital contribution arising from lattice-expansion-induced changes in the band structure.
The electronic specific heat, the magnetic susceptibility, and the $T^2$ coefficient in resistivity point to a weakly-correlated 3D Fermi-liquid state with a modest electron correlation, as supported by the Wilson and Kadowaki-Woods ratios.
\end{abstract}
\maketitle

\section{Introduction}

Ruthenium dioxide RuO$_2$ in the rutile structure has provided a wide variety of scientific as well as technological interests.
It is known as a Pauli paramagnetic metal since 1930s \cite{guthrie1931magnetic,ryden1970magnetic, ryden1970electrical, passenheim1969heat,fletcher1968magnetic}. 
However, fitting of the specific heat with a conventional $T^3$ term had lead to an unusually high Debye temperature \cite{passenheim1969heat, mertig1986specific} due to low-frequency optical phonons with frequencies overlapping with acoustic phonons. 
With glassy impurities, it is used as a thermometer with small magnetoresistance at low temperatures \cite{Doi_RuO2-sensor_LT17-1984, Myers2021}. 
It is also well known as a catalyst for the oxygen evolution reaction (OER)~\cite{Over2000catalyst, Ping2024catalyst}. Recently, superconductivity was induced under anisotropic strain \cite{ruf2021strain} and reproduced well in the following study \cite{uchida2020superconductivity}. 

In 2017, an antiferromagnetic (AFM) ordering with a small room-temperature magnetic moment of approximately 0.05$\mu_\textrm{B}$ ($\mu_\textrm{B}$ is the Bohr magneton) was reported from polarized neutron diffraction \cite{berlijn2017itinerant}.
The AFM ordering with the N\`eel temperature $T_\textrm{N}>$ 300 K was confirmed with resonant x-ray scattering \cite{zhu2019anomalous} and angle-resolved photoemission spectroscopy (ARPES)~\cite{Lin2024ARPES}. 
However, more recent studies with $\mu$SR \cite{hiraishi2024nonmagnetic}, neutron diffraction \cite{Kiefer2025}, as well as ARPES \cite{osumi2025ARPES} did not detect evidence for AFM ordered state.

The issue of intrinsic magnetic properties is particularly important since RuO$_2$ is considered among the most promising candidate materials of altermagnetism with a large exchange-splitting energy \cite{vsmejkal2022beyond, vsmejkal2022emerging}.
This novel magnetic state in a certain class of AFM was studied previously \cite{Noda2016} and more recent studies of spin splitting, spin-current generation, and anomalous Hall effect \cite{Ahn2019, Naka2019, vsmejkal2020crystal, Hayami2019} prepared the stage for the recognition as altermagnetism.
The observed anomalous zero-field Hall effect and spin-torque effect in thin films of RuO$_2$ are interpreted in terms of altermagnetism \cite{karube2022observation,feng2022anomalous}.
Using ultra-clean crystals \cite{paul2025growth}, we demonstrated by ARPES \cite{osumi2025ARPES} and quantum oscillations\cite{wu2025fermi} that bulk RuO$_2$ is paramagnetic down to low temperatures.

Nevertheless, bulk characterization of the paramagnetic state of RuO$_2$ in terms of a Fermi liquid has not been systematically provided using these ultra-clean crystals.
We report here the Fermi-liquid universality of RuO$_2$ from resistivity, specific heat and magnetic susceptibility. 
We found that it exhibits a temperature-dependent susceptibility that cannot be explained by the energy-dependent density of states (DOS) with thermal activation of quasiparticles, but is attributable to orbital contribution.

\section{Experimental Methods}
Single crystals of RuO$_2$ were grown using the sublimation transport method \cite{paul2025growth,huang1982growth}. 
High-purity RuO$_2$ powder (99.95$\%$, Rare Metallic Ltd.) was placed in an alumina crucible and heated to 1250$^\circ$C under a continuous oxygen flow at 50 cm$^3$/min and the heater was turned off after 100 hrs. 
Single crystals formed in the lower temperature region at around 1000$^\circ$C and predominantly exhibit the (101) facets. 
The typical crystal dimensions were approximately \(5 \times 3 \times 1\sim2 \ \mathrm{mm}^3\) as shown in Fig. \ref{fig1} (b). 
The crystal planes were identified using x-ray Laue images as well as diffraction peaks of x-ray diffractometer (XRD). 
Powder x-ray diffraction was used to confirm the phase purity of the crystals. 
The lattice parameters at room temperature were determined as $a$ = 4.49 \AA \ and $c$ = 3.11 \AA \cite{paul2025growth}, consistent with previously reported values \cite{ryden1970electrical}. 

For resistivity measurements, we used MPMS-XL (Quantum Design) with a self-designed transport probe.
AC four-probe measurements were performed using a lock-in amplifier (Stanford Research Systems, SR830) with its internal oscillator.
A low frequency of 17 Hz was chosen to suppress phase shifts, and an excitation current of 10 mA (rms) was used.
The measurement technique is described in more detail in \cite{paul2025growth}.
Specific heat was measured using a commercial calorimeter system with a helium-4 refrigerator (Quantum Design, PPMS).
The thermometer was calibrated for each magnetic field used in this study. 
Magnetization was measured using a superconducting quantum interference device (SQUID) magnetometer (Quantum Design, MPMS-XL), 

\section{FERMI-LIQUID CHARACTERISTICS}

\subsection{Resistivity}
\begin{figure}
		\begin{center}
			\includegraphics[scale=0.29]{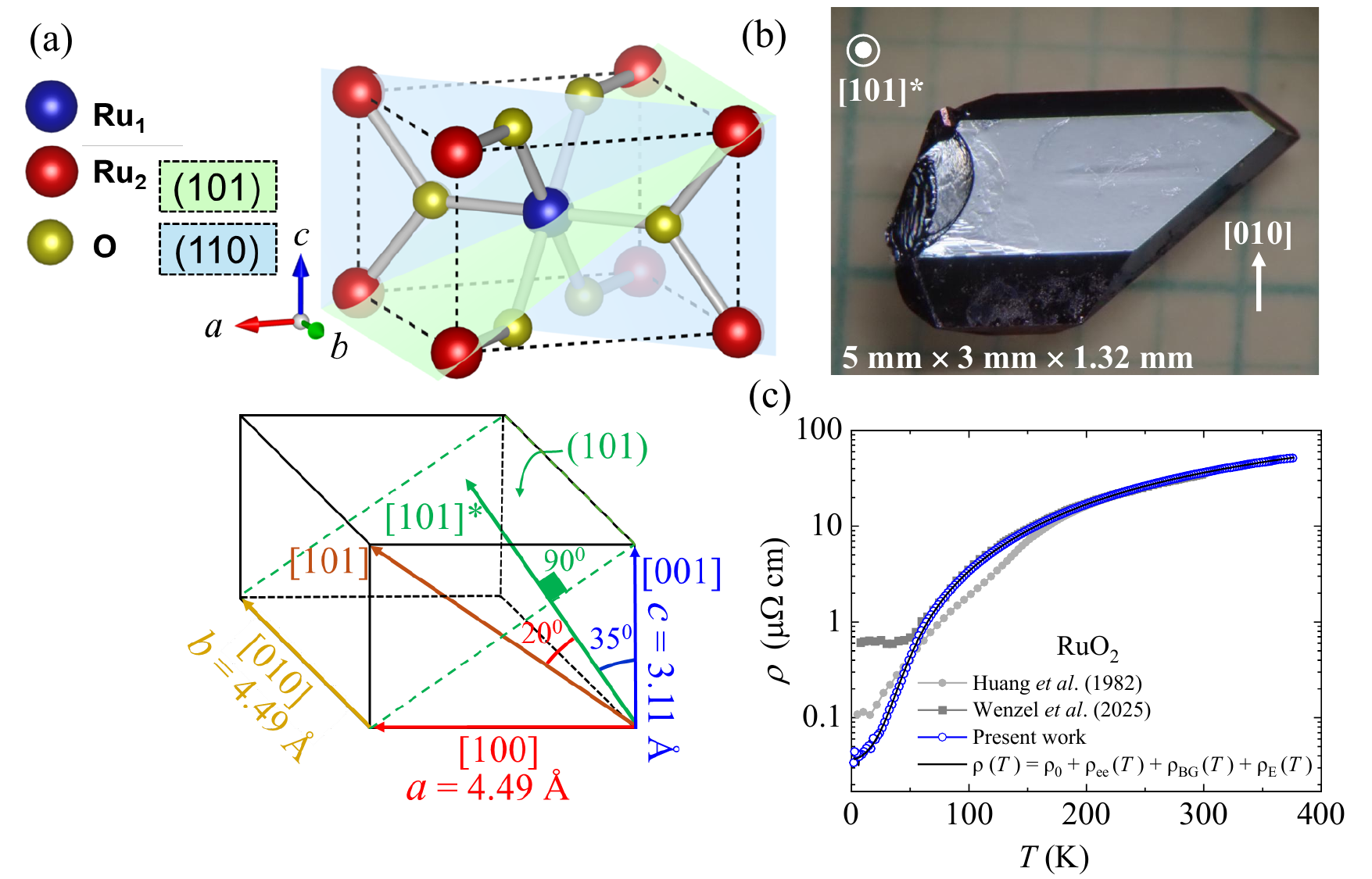}
		\caption{Characterization of RuO$_2$ single crystal: (a) Schematic of the rutile crystal structure \cite{momma2008vesta}. In the anticipated antiferromagnetic phase, the Ru sites shown in red (blue) would have magnetic moments up (down) along the [001] direction. The bottom figure shows that the [101]* direction, perpendicular the (101) plane, is about 20 degrees different from the [101] direction. (b) Optical image of a typical RuO$_2$ single crystal with the size 5$\times$3$\times$1.3 mm$^3$. Crystal orientation is confirmed by x-ray Laue photos. (c) Resistivity with a current along the [001] axis. Needle-shaped crystal elongated along the [001] direction is used \cite{paul2025growth}. The residual resistivity ratio (RRR) is as large as 1200. The current directions is along the [100] axis for Wenzel $et$ $al.$\cite{wenzel2025fermi}; but not specified in Huang $et\ al.$ \cite{huang1982growth}. }
	\label{fig1}
	\end{center}
\end{figure}

Figure \ref{fig1} (c) represents the resistivity with the current along the [001] direction, plotted against temperature.
The details are described in \cite{paul2025growth}, but here we plotted the resistivity using different voltage leads (V3-V4) from those presented in \cite{paul2025growth}.
The room-temperature resistivity of 36 $\mu\Omega$-cm is consistent with previous reports \cite{huang1982growth,pawula2024multiband,wenzel2025fermi}.
The residual resistivity of 37 n$\Omega$cm corresponds to the mean free path of $\sim$ 0.44 $\mathrm{\mu}$m.
The residual resistivity ratio RRR = $\rho_{300 \mathrm{K}}$/$\rho_{2\mathrm{K}}$ of 1200 is the highest among those reports, indicating the high quality of the single crystals used in this study. For the present specific heat and magnetization studies, single crystals from the same batch as the ones with RRR up to 1200 were used. 
We then carried out the fitting for the resistivity over the temperature range from 2 to 375 K using Bloch-Grüneisen (BG) and Einstein models. 
The obtained Debye temperature $\theta_\mathrm{D}$ = 406 K and Einstein temperature $\theta_\mathrm{E}$ = 835 K \cite{paul2025growth} agree well with those reported in Ref. \cite{lin2004low}.

\subsection{Specific heat}
\begin{figure*}[t]
		\begin{center} 
		\includegraphics[scale=0.54]{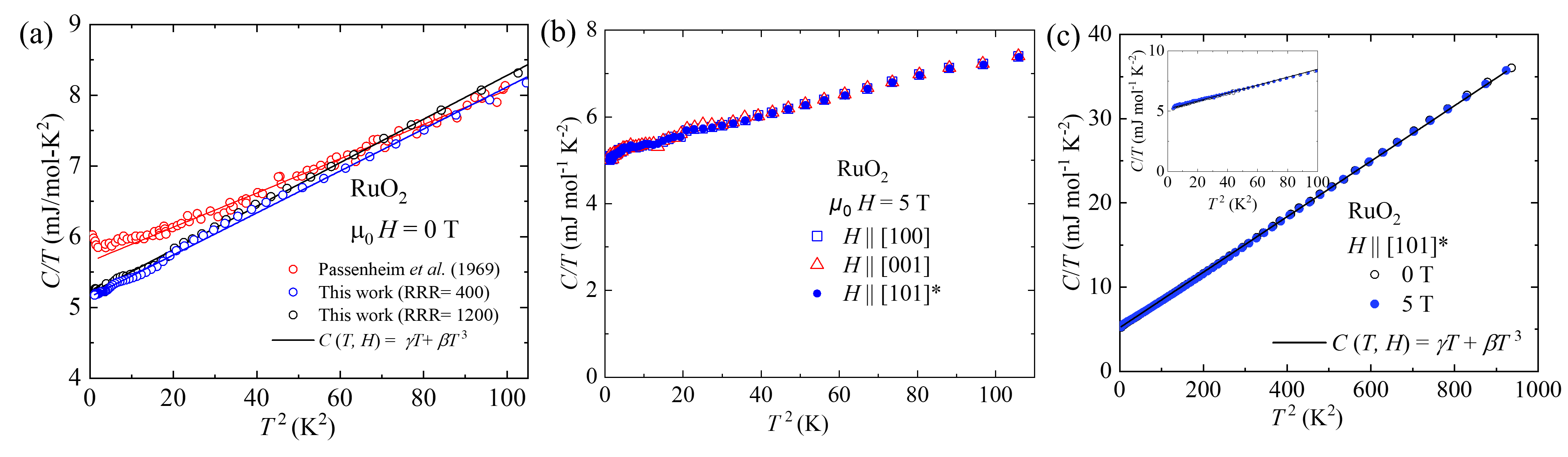}
		\caption{(a) Variations of the zero-field specific heat divided by temperature, $C/T$, of a RuO$_2$ single crystal measured from 1.1 to 10 K, plotted against $T^2$. Fitted curves with Eq. (\ref{eq3}) are also shown. The results of the present studies (blue and black circles) and those of Passenheim and McCollum \cite{passenheim1969heat} (red circles) are shown for comparison. (b) Specific heat of RuO$_2$ divided by temperature plotted against $T^2$, measured from 1.1 to 10 K under the magnetic field of 5 T applied along the [101]*, [001], and [010] directions. It shows no observable anisotropy. (c) $C/T$ vs $T^2$ plotted for 0 T and 5 T along the [101]* direction for the sample with RRR = 1200. The inset shows the data up to $T$ = 10 K, indicating only a small field dependence.}
	\label{fig2}
	\end{center}
\end{figure*}

Figure \ref{fig2} (a) presents the temperature dependence of the specific heat of RuO$_2$ single crystals with the RRR of 400 and 1200 at zero field, compared with a previous report. We fit with a standard $T$-linear electronic term and a Debye-$T^3$ phonon term:
\begin{equation}
    C (T) = \gamma\,  T + \beta\, T^3.
    \label{eq3}
    \end{equation}
\noindent Here $\gamma$ is the Sommerfeld coefficient, which in the free-electron model, is given by:
\begin{equation}
     \gamma= \frac{\pi^2 k_\mathrm{B}^2 N_\mathrm{A} D(\epsilon_\mathrm{F})}{3},
     \label{eq4}
    \end{equation}
\noindent where $k_\mathrm{B}$ is Boltzmann's constant, $N_\textrm{A}$ is Avogadro's number, and $D(\epsilon_\mathrm{F})$ is the density of states (DOS) at the Fermi energy that includes both spin directions.
If the second term in Eq. \eqref{eq3} contains purely phonon contribution with the Debye temperature $\theta_\mathrm{D}$, $\beta = \beta_\mathrm{ph}$ given by the following:
\begin{equation}
     \beta_\textrm{ph}=  \frac{12 \pi^4 k_\mathrm{B} N_\mathrm{A} N} {5\ \theta_\mathrm{D}^3},
      \label{eq5}
    \end{equation}
\noindent where $N$ = 3 is the number of atoms in a formula unit for RuO$_2$.
\begin{table*}
 	\caption{Comparison of the fitting parameters for specific heat of RuO$_2$. The fitting parameters are derived using Eq.(\ref{eq3}).}
 	\label{tab1}
 	\begin{tabular*}{0.8\textwidth}{@{\extracolsep{\fill}}cccccccccl}
 		\hline
 		\hline
 		\   parameters &  single crystal  &single crystal& single crystal \cite{mertig1986specific} & polycrystals \cite{passenheim1969heat}      \\
    &  (this work) & (this work) &  &       \\
 		\hline
        RRR & 400 & 1200  & 500 & --   \\
 		\ sample mass & 36.4 mg & 48.1 mg  & 59.1 mg & 18.74 g   \\
          $\gamma$ (mJ mol$^{-1}$K$^{-2}$) & 5.18  & 5.12 & 5.20   & 5.77    \\
         $\beta$ (mJ mol$^{-1}$K$^{-4}$) & 0.0289&0.0317 & 0.0376  & 0.0225  \\
          $\theta_\mathrm{D}$ (K) & 587 & 569  & 544  & 637    \\
          $D$($\epsilon_\mathrm{F}$) (cell spin Ryd)$^{-1}$ &29.2 & 29.5  & 30.0  & 33.2 \\
		\hline
		\hline
	\end{tabular*}
\end{table*}
The measured specific heat is in reasonable agreement with the literature for a polycrystalline sample down to 0.5 K \cite{passenheim1969heat} as well as for a single crystalline sample measured down to 2 K \cite{mertig1986specific}.
Nevertheless, the nuclear quadrupole contribution reported in Ref. \cite{passenheim1969heat} is absent in our data, consistent with a small quadrupolar splitting of a mK scale in the nuclear magnetic resonance (NMR) \cite{mukuda1999spin}. 
 Fitting our zero-field data of RRR=400 crystal using Eq. (\ref{eq3}) yields $\gamma$ = 5.18 mJ mol$^{-1}$K$^{-2}$, $\beta$ = 0.0289 mJ mol$^{-1}$K$^{-4}$, corresponding to $\theta_\mathrm{D}$ = 587 K. 
This is noticeably higher than $\theta_\mathrm{D} \sim$406 K from the BG fitting of the resistivity.
In RuO$_2$, a total of 18 vibrational modes are present. Two low-frequency optical phonons have energies comparable to the Debye temperature \cite{Bohnen2007phonons}. 
Thus, $\beta$ in Eq. (\ref{eq5}) properly contains both acoustic and optical phonon excitations at low temperatures.

Figure \ref{fig2} (b) illustrates the direction-dependent specific heat, measured in the magnetic field of 5 T. To investigate the contribution of spin fluctuations, we applied the magnetic field perpendicular to three crystallographic planes: (100), (001), and (101). In particular, no anisotropy was observed at a field strength of 5 T.  
The specific heat was measured in magnetic fields of 0 T and 5 T along the [101]$^*$ axis, up to 30 K, as shown in Fig. \ref{fig2} (c).
No significant field dependence is observed up to 5 T, consistent with the expected paramagnetic Fermi-liquid behavior.

We compare the value of $\gamma$ with $D(\epsilon_{\textrm{F}})$ to examine the mass enhancement.
From $\gamma$ and Eq. (\ref{eq4}), $D(\epsilon_{\textrm{F}})$ = 4.29 states/eV/f.u. = 29.2 states/Ry/spin. 
These value is somewhat higher than 3.834 states/eV/f.u. = 26.07 states/Ry/spin/f.u. from the band-structure calculation without spin-orbit coupling (SOC) and 4.067 states/eV/f.u. = 27.67 states/Ry/spin/f.u. including SOC \cite{osumi2025ARPES}. 
Here f.u. stands for formula unit.
The on-site Coulomb repulsion $U$ is set to zero in these calculations.
The mass enhancement factors are given as follows \cite{Vollhardt-Woelfle1990}: 
\begin{equation}
\begin{split}
D(\epsilon_\mathrm{F}) &= D_\mathrm{{band}} (1+ \lambda)= D_\textrm{FG}\ . \frac{m^*}{m} \\
                & = D_\textrm{FG} \left(1+ \frac{F^\textrm{s}_1}{3}\right),
\end{split}
\label{eq7}
\end{equation}
\noindent where $D_\textrm{FG}$ is DOS at $\epsilon_\textrm{F}$ for a Fermi gas, 0.662 states/eV/f.u. for RuO$_2$, and $m^*$ is the effective mass of the electrons. 
We obtain the enhancement factor $\lambda$ = 0.055 using the calculated value including SOC, and the Landau Fermi-liquid parameter $F^\textrm{s}_1$ = 16.5 from $\gamma$.
Thus, the mass enhancement $m^*/m \approx 6.49$ is substantial, but further enhancement over the band mass, $\lambda$, is small.  

\begin{figure*}
    \begin{center}
    \includegraphics[scale=0.7]{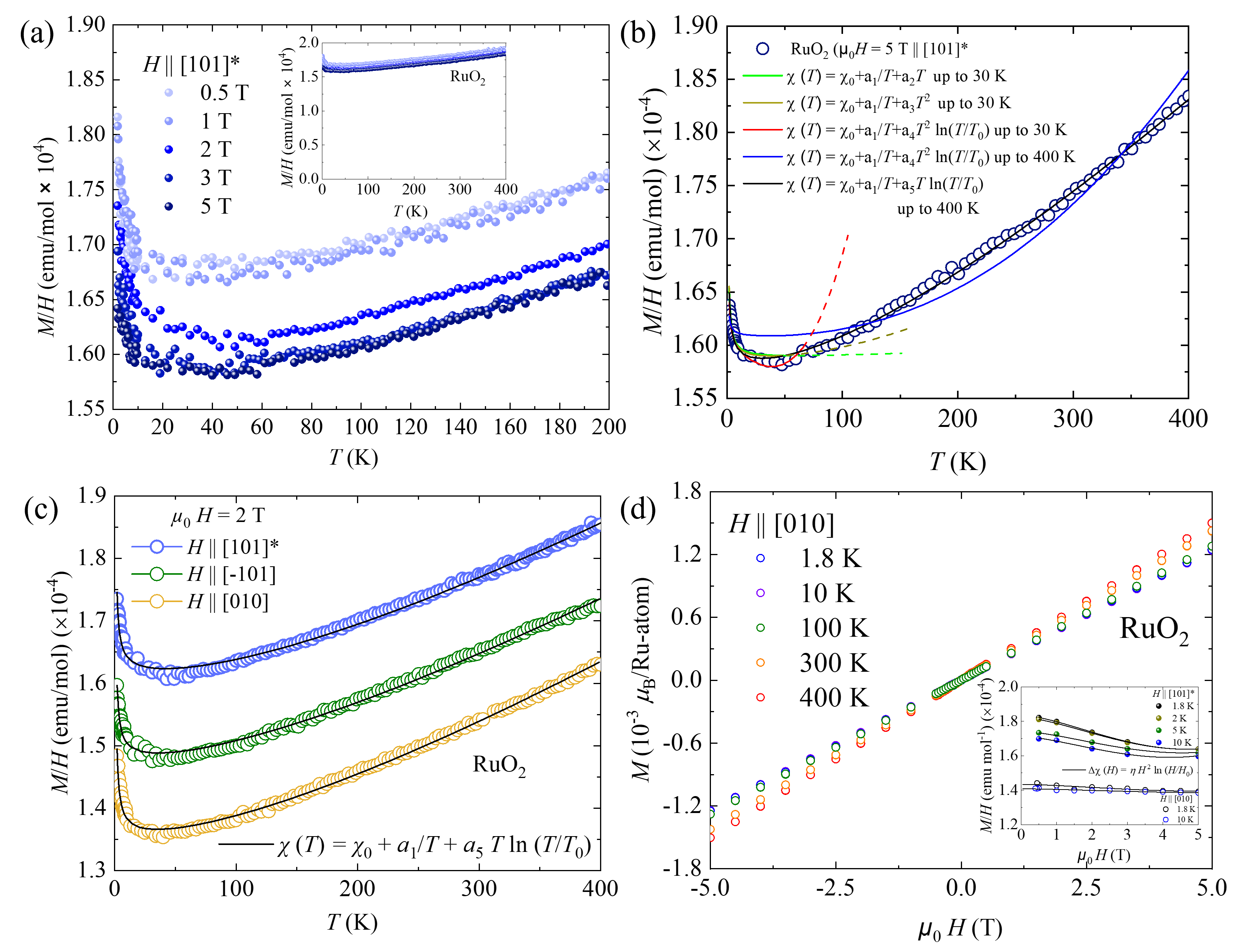}
    \caption{(a) Variation of the DC magnetic susceptibility, $M/H$, of RuO$_2$ with temperature from 1.8 K to 200 K under various magnetic fields $\mu_{0}H$ applied perpendicular to the (101) plane. The inset shows the susceptibility over a broader temperature range (1.8-400 K), revealing no signs of a magnetic transition. (b) Comparison of different fitting functions for the DC susceptibility of RuO$_2$ at 5 T. Equations (\ref{eq12}), (\ref{eq13}), and (\ref{eq14}) give reasonable fitting only up to 30 K, but deviate from the data substantially at higher temperatures. In contrast, a phenomenological form Eq. (\ref{eq14a}) fits the data well all the way up to 400 K. (c) Anisotropy of $M/H$ at 2 T along different field directions, [101]*, [010], and [-101]. The susceptibility is well fitted with $T\ \mathrm{ln} (T/T_0)$ over a wide temperature range. (d) Magnetization vs magnetic field along the [010] direction at various temperatures. Inset shows variation of the DC susceptibility with field at various temperatures up to 10 K. The susceptibility is fitted with $\Delta\chi (H)$ = $\eta$ $H^2\mathrm{ln}(H/H_0)$ (Eq. \ref{eq16b}). The data for $H \parallel [010]$ is taken from $M$ vs $H$, while the data for $H \parallel [101]^*$ is taken from $M/H$ vs $T$ in Fig. \ref{fig3} (a).}
    \label{fig3}
    \end{center}
\end{figure*}
\begin{table*}
    \small
    \caption{Comparison of the Fermi-liquid and other parameters of RuO$_2$ in this work (RRR = 400). Here, the density of states D is expressed in units of states eV$^{-1}$\ f.u.$^{-1}$.}
 	\label{tab2}
    \begin{tabular*}{0.9\textwidth}{@{\extracolsep{\fill}}lllllllllll}
 	\hline \hline
 	\ &  parameters &  RuO$_2$ (this work)     \\ \hline
 	\ Effective mass ratio & $m^* / m$ & 6.265  \\
        \ Landau Fermi-liquid &&&\\
      parameters  & $F\mathrm{_1^s}$ & 15.8      \\
        \   & $F\mathrm{_0^a}$ & $-$0.867     
        \\ \hline
         \ RRR  &$\rho_{300\ \mathrm{K}}/\rho_{2\mathrm{K}}$    \\
        \ Carrier density  &$n$ ($\mathrm{m}^{-3}$) &  0.128$\times 10 ^{30}$     \\
        \ Sommerfield coefficient & $\gamma$ (mJ mol$^{-1}$K$^{-2}$) & 5.18     \\
        \ Phonon $T^3$ contribution & $\beta_\mathrm{ph}$ (mJ mol$^{-1}$K$^{-4}$) & 0.0289   \\
        \ Coupling constant ratio &$\lambda$ &  0.055 \\
        \ Characteristic temperature  & $T_0$ (K) & 30     \\
        \ Characteristic field & $\mu_0 H_0$ (T) & 8.5\  ($H\|$ [010])     \\
        \ Residual resistivity & $\rho_0$ ($\mu\Omega \ \mathrm{cm}$) & 0.037      \\
        \ Fermi energy (Fermi gas) & $\epsilon_\mathrm{F}$ (eV) & 9.243     \\
        \ Density of states (Fermi gas)  & $D_0 (\epsilon_\mathrm{F})$ (states eV$^{-1}$\ f.u.$^{-1}$) & 0.649    \\
        \ Density of states (band calc.) & $D_{\mathrm{band}}$ (states eV$^{-1}$\ f.u.$^{-1}$) &  4.067      \\
        \ Density of states ($\gamma$ in $C_P$)  &$D_{\mathrm{exp}}$ (states eV$^{-1}$\ f.u.$^{-1}$) &  4.29      \\
        \hline \hline
    \end{tabular*}
\end{table*}
\subsection{Magnetization}
Figure \ref{fig3} (a) represents the DC susceptibility, the magnetization $M$ divided by $H$, as a function of temperature up to 200 K under various magnetic fields applied along the [101]* direction. 
The inset illustrates the $M/H$ vs $T$ up to 400 K; no signature of a magnetic phase transition is observed.
We include the low-temperature upturn in magnetization below 30 K using the Curie law term $a_1$/$T$ in Eq. \ref{eq12}, in which $\chi_0$ and $a_2(H)T$ represent, respectively, the temperature-independent and temperature-field dependent contributions in the Pauli susceptibility:
\begin{equation}
    \chi \left(T,H\right) = \chi _0 + \frac{a_1}{T} + a_2(H)\ T.
    \label{eq12}
    \end{equation}
For a spin-1 impurity with concentration $x$, such as a localized Ru$^{4+}$ ion, the expected Curie term is given as $a_1= (g \mu_\mathrm{B})^2 S(S+1)x/3k_\mathrm{B}$= 0.5$x$ emu-K/mol.  
From the fitting of the data for RRR = 400 with Eq. (\ref{eq12}), we obtain $a_1$= 1.8$\times 10^{-5}$ emu-K/mol, which corresponds to the impurity concentration of $x$ = 36 ppm \cite{paul2025growth}.
This gives the upper limit of the paramagnetic-impurity concentration.
For comparism, Curie terms in low temperature magnetization reported from different groups somewhat differ: the ratios $M$(2 K)/$M$(20 K) (and the RRR values) are 1.08 (400) and 1.01 (1200) in the present work, 1.13 (158) in Ref. \cite{Kiefer2025}, and 1.84 (118) in Ref. \cite{wenzel2025fermi}. 

For another source of temperature dependence, the DC susceptibility weakly decreases with decreasing temperature from 400 K, as shown in the inset of Fig. \ref{fig3} (a). 
To express such behaviour, we introduce a fitting with \cref{eq12} as shown in Fig. \ref{fig3} (b).
Although the fitting is satisfactory below 30 K, it deviates significantly at higher temperatures.
Adding various higher-order polynomial terms in $T$ (such as Eq. \eqref{eq13}) also does not improve the fitting. 

\begin{equation}
    \chi \left(T,H\right) = \chi_0 + \frac{a_1}{T} + a_3(H)\ T^2.
    \label{eq13} 
    \end{equation}

As an alternative approach, we introduce a $T^2 \textrm{ln} T$ term given in Eq. (\ref{eq14}), commonly used for spin fluctuation systems \cite{DeVisser1987}. 
\begin{equation}
    \chi \left(T,H\right) = \chi_0 + \frac{a_1}{T} + a_4(H)\ T^2\  \mathrm{ln}\left(\frac{T}{T_0 (H)}\right).
    \label{eq14}
    \end{equation}
\noindent As shown in Fig. \ref{fig3} (b), Eq. (\ref{eq14}) cannot represent the observed behaviour in a wider temperature range up to 400 K.
Thus, we introduce Eq. (\ref{eq14a}) with another logarithmic temperature-dependent susceptibility, which fits the data very well up to 400 K (Figs. \ref{fig3} (b) and (c)).
\begin{equation}
    \chi \left(T,H\right) = \chi_0 + \frac{a_1}{T} + a_5(H)\ T\  \mathrm{ln}\left(\frac{T}{T_0 (H)}\right).
    \label{eq14a}
\end{equation}
\noindent The zero-temperature Pauli susceptibility of RuO$_2$, obtained from Eq. (\ref{eq14a}) at 5 T along the $ [101]^*$ direction, is $\chi_0$=1.58$\times$10$^{-4}$ emu/mol = 1.984 $\times 10^{-9}$ m$^3$/mol = 5.77 $\times 10^{-5}$ in the electromagnetic unit (EMU) and SI units, which are consistent with previous reports \cite{berlijn2017itinerant, wenzel2025fermi}.
For the logarithmic term, we obtain $a_5$ = 1.74$\times10^{-8}$ and 2.38$\times10^{-8}$  emu/K$^2$-mol at 0.5 and 5 T, respectively, while the characteristic temperature $T_0=30$ K remains unchanged.
According to Eq. (\ref{eq14a}), the logarithmic contribution gives the minimum in $\chi$ at $T_0/e$.

Let us first examine if Eq. (\ref{eq14a}) phenomenologically represents the energy dependence of DOS near $\epsilon_\mathrm{F}$.
The DOS $D(\epsilon)$ for the correlation energy $U = 0$ that gives a non-magnetic state is convex upward near $\epsilon_\mathrm{F}$, as shown in appendix \ref{A1}(a). 
Considering this shape alone, the thermal broadening of the Fermi distribution would lead to a decrease in the average DOS with increasing temperature. 
For a nonmagnetic system, the spin susceptibility is given by
\begin{equation}
\chi(T) = \mu_\mathrm{B}^2 \int_0^\infty D(\epsilon) \left( -\frac{\partial f(\epsilon, \mu, T)}{\partial \epsilon} \right) d\epsilon,
\label{eqa0}
\end{equation}
where $f$ is the Fermi--Dirac distribution function.
This expression indicates that $\chi(T)$ is proportional to the thermal average of the DOS near the Fermi level.

We evaluated $\chi(T)$ by directly performing the numerical integration of \cref{eqa0} and obtained results shown in appendix Fig. \ref{A1}(b).
$\chi(T)$ decreases more strongly at higher temperatures with an overall reduction of about $5\%$ between $10$ K and $300$ K.
This is in sharp contrast with the experimentally observed $increase$ by 10\%.
Thus, a simple thermal activation picture based solely on the DOS is insufficient to explain the experimentally observed anomalous positive temperature coefficient.

Similar positive susceptibility coefficients have been reported in transition metals such as Ti \cite{mori1965orbital,collins1971some}.
In these transition metals, total susceptibility consists of $\chi_\mathrm{tot} =  \chi_\mathrm{s} + \chi_\mathrm{orb}+\chi_\mathrm{dia}$, where the first term is the Pauli susceptibility, the second term the orbital ``Van Vleck" susceptibility, and the third term the core diamagnetic susceptibility.  
Pure Ti shows magnetic anisotropy in different field directions of about 20\% and a change in susceptibility with temperature of about 5\% over 4.2–300 K, while the lattice thermal expansion is only about 0.18\% \cite{collins1971some}.
Such characteristic behavior has been attributed to changes in the orbital Van Vleck energy gap induced by lattice thermal expansion \cite{collins1971some}.
The magnitudes of temperature dependence and anisotropy of magnetic susceptibility, as well as of the lattice expansion, are comparable to those in RuO$_2$.
From \cref{fig3} (c), susceptibility change is about 10\% with a magnetic anisotropy of about 15\% over 10–300 K.
It is likely that orbital contribution plays an important role in such a characteristic susceptibility.
Ryden $et\  al.$ \cite{ryden1970magnetic} estimated $\chi_\mathrm{orb}$ to be approximately 30$\times10^{-6}$ emu/mol and $\chi_\mathrm{dia}$ to be $-15\times10^{-6}$ emu/mol, giving a temperature-independent susceptibility $\chi_0$ = $\chi_\mathrm{s} + 15\times10^{-6}$emu/mol.
Concerning the temperature dependence, the lattice parameter $a$ increases by 0.07\%, whereas $c$  decreases by 0.06\% from 80 to 250 K \cite{Kiefer2025}. 
Based on these overall similarities, the characteristic temperature dependence of the susceptibility in RuO$_2$ is attributable to the lattice-expansion–induced variation of orbital paramagnetism.

\begin{figure*}
	\begin{center}
			\includegraphics[scale=0.7]{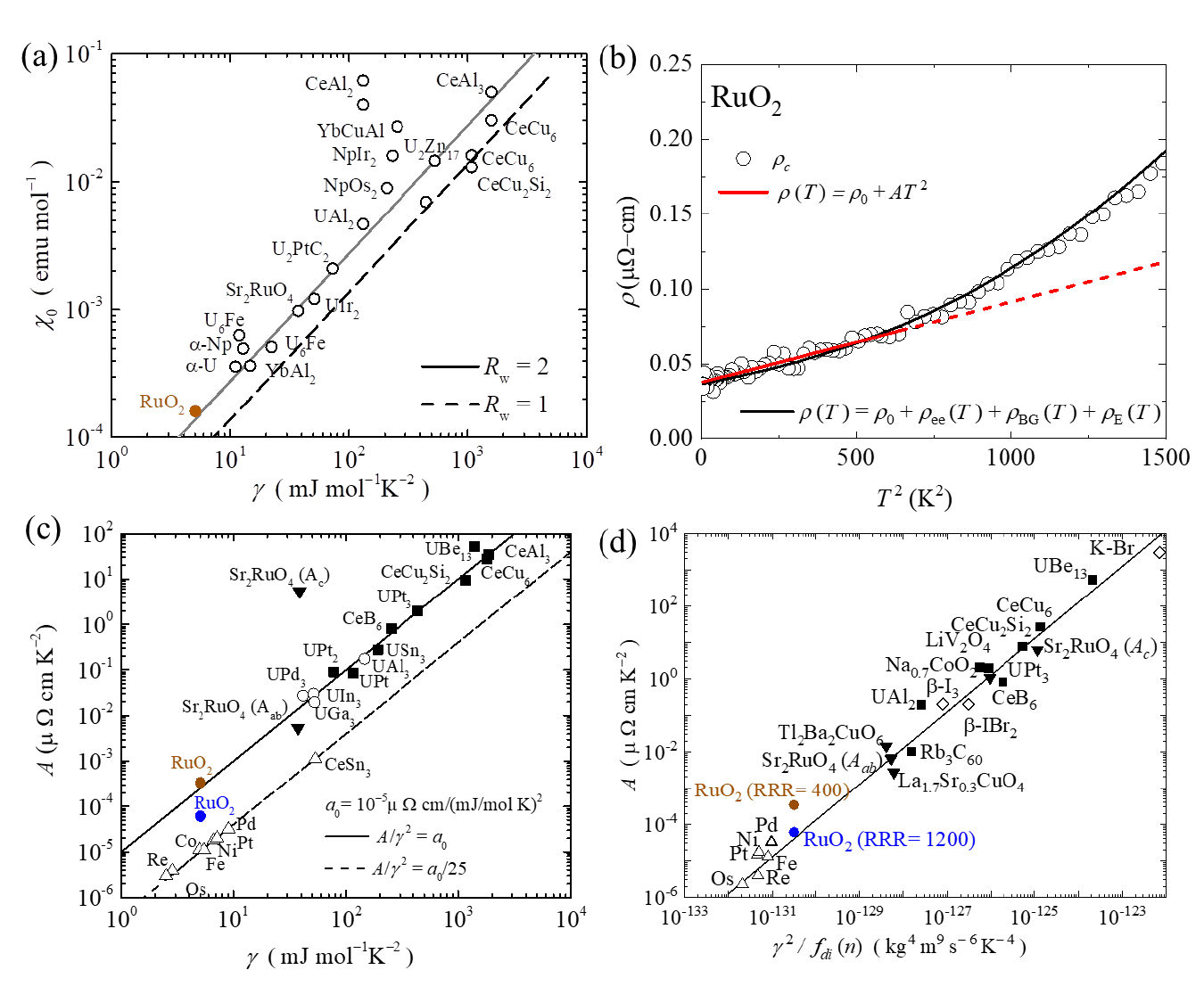}
		\caption{(a) Pauli susceptibility $\chi_0$ in Eq. \eqref{eq16} plotted on a logarithmic scale against the Sommerfeld coefficient $\gamma$ in the specific heat for various compounds including RuO$_2$. 
        The solid and dashed lines represent the Wilson ratio $R_\mathrm{W}$ of 2 and 1, respectively. 
        (b) The electrical resistivity of RuO$_2$ for current along the [001] direction plotted against $T^2$. 
        The solid line represents the fitting below 25 K using $\rho (T)= \rho_0 + AT^2$ with $A$ = 0.053 n$\Omega$-cm/K$^2$. 
        (c) The $T^2$ resistivity coefficient $A$ plotted on a logarithmic scale against $\gamma$. 
        Solid and dashed lines represents the Kadowaki–Woods ratio (KWR) $A/\gamma^2 = a_0$=1.0$\times$10$^{-5}$ $\mu \Omega$-cm/(mJ/mol-K)$^2$ and a$_0$/25, respectively \cite{kadowaki1986universal,miyake1989relation,maeno1997two,hussey2005,jacko2009unified}. (d) Plot of the Kadowaki–Woods-Jacko ratio ($R_\mathrm{{KWJR}}$) given by Eq. \eqref{eq18} \cite{jacko2009unified} including RuO$_2$.}  
	\label{fig4}
        \end{center}
\end{figure*}

A weak magnetic anisotropy of about 15$\%$ is observed when the magnetic field is applied along different crystalline directions as shown in Fig. \ref{fig3} (c): with 2-T field along the [101]*, [$-$101] and [010] directions at 2 K, $\chi$$_{[101]^*}$= 1.75, $\chi$$_{[-101]}$=1.55, and $\chi$$_{[010]}$=1.50 $\times$10$^{-4}$ emu/mol, respectively. 
This anisotropic behavior is consistent with previous results \cite{berlijn2017itinerant, Kiefer2025, ryden1970electrical}, but in contrast with isotropic response of the specific heat under fields shown in Fig. \ref{fig2}.
The nearly parallel shift for different field directions suggests that the anisotropy mainly comes from the $\chi_0$ term.
Considering the highly isotropic specific heat under field, such modest anisotropy in $\chi_0$ is attributed to the anisotropy of the orbital contribution.

The magnetization curves at various temperatures from 1.8 K to 400 K are shown in Fig. \ref{fig3} (d). 
These $M$-$H$ curves exhibit no hysteresis, consistent with a paramagnetic state.
The slope $M/H$ is temperature and field-dependent, decreasing with increasing field, as shown in the inset of Fig. \ref{fig3} (d).
Based on the experimentally observed field dependence, we introduce a phenomenological term $\Delta \chi (H)$:
\begin{equation}
\begin{split}
    \chi \left(T,H\right) &= \chi_0 + \frac{a_1}{T} + a_5(H=0)\ T\  \mathrm{ln}\left(\frac{T}{T_0 (H=0)}\right)\\+
                          &\Delta \chi (H),
    \end{split}
    \label{eq16}
    \end{equation}
where 
\begin{equation}
\begin{split}
   \Delta \chi (H) = a_6 H^2 \mathrm{ln} \left(\frac{H}{H_0}\right).
    \end{split}
    \label{eq16b}
    \end{equation}
\noindent As shown in the inset of Fig. \ref{fig3} (d), the characteristic field obtained from this fitting is $\mu_0 H_0(T)$ = 7.7 T for $H\parallel[101]^*$ and 8.5 T for $H\parallel[010]$.

The Fermi-liquid parameter $F^\textrm{a}_0$ is defined as follows \cite{Vollhardt-Woelfle1990}:
\begin{equation}
\chi_0 = \frac{\chi_\textrm{FG}}{1 + F^\textrm{a}_0}, \ \ \chi_\textrm{FG}= \mu_0 \mu_\textrm{B}^2 D_{\textrm{FG}}.
\label{eq15}
\end{equation}
Using the Pauli susceptibility for the Fermi-gas model $\chi_\textrm{FG} = 0.878 \times 10^{-5}$, we obtain $F^\textrm{a}_0 = -0.867$.
Table \ref{tab2} summarises the Landau Fermi liquid parameters $F^\textrm{s}_1$, $F^\textrm{a}_0$ and other parameters of RuO$_2$.

\subsection{Universality plots}
To characterize the strength of electronic correlations, the Wilson ratio 
\begin{equation}
R_\mathrm{W}= \frac{4 \pi^2 k_\textrm{B}^2}{3\mu_{0} g^2 \mu_\textrm{B}^2}\left(\frac{\chi_0}{\gamma}\right)
 \label{eq17}
\end{equation}

\noindent is widely used.
Although $\chi_0$ contains both positive orbital Van Vleck susceptibility and negative diamagnetic contributions, these compensating contributions are not large compared with $\chi_\mathrm{s}$; therefore, in the following, $\chi_0$ is taken to represent the Pauli susceptibility.
For a free electron gas, $R_\mathrm{W}$ is unity, and it increases up to 2 for strongly correlated materials \cite{Engelbrecht1995Wilson-ratio}. 
Figure \ref{fig4} (a) presents $\chi_0$ versus $\gamma$ for RuO$_2$ along with other strongly and weakly correlated electron systems. 
The Wilson ratio for RuO$_2$ using $\chi_0$ is found to be 2.3. 
RuO$_2$ may not be considered as a strongly correlated material, considering the modest mass enhancement reflected in the small Sommerfeld coefficient $\gamma$ = 5.18 mJ mol$^{-1}$K$^{-2}$, compared with 40 mJ mol$^{-1}$K$^{-2}$ for Sr$_2$RuO$_4$ with a similar carrier density, and in the correspondingly small effective mass obtained from quantum oscillations \cite{wu2025fermi}.
Magnetization exhibits some anisotropy (Fig. \ref{fig3} (c)), in contrast to the isotropic specific heat (Fig. \ref{fig2} (b));
such magnetic anisotropy suggests additional magnetic contribution that enhances the Wilson ratio.
Recent plasma frequency studies also classify RuO$_2$ as a weakly correlated paramagnetic metal \cite{wenzel2025fermi}.

Another key parameter for universal characterization of quantum materials is the Kadowaki–Woods ratio (KWR) \cite{kadowaki1986universal}, defined as $R_\mathrm{{KW}}= A$/$\gamma^2$, where $A$ is the coefficient of the $T^2$ term in resistivity.
Strongly-correlated materials empirically follow $R_\mathrm{{KW}}=a_0$= 10$^{-5}$ $\mu \Omega$ cm/(mJ/mol K)$^2$.  
The resistivity of RuO$_2$ at low temperatures follows $\rho$($T$)=$\rho_0$ + $A$$T^2$, as shown in Fig. \ref{fig4} (b).
The coefficient $A$, estimated by fitting the resistivity data below 25 K, is 0.053 n$\Omega$-cm/$\mathrm{K}^2$.
This value is two orders of magnitude lower than that of Sr$_2$RuO$_4$ ($A$ = 4.5-7.5 n$\Omega$-cm/$\mathrm{K}^2$) \cite{maeno1997two}. 
\color{black}
Figure \ref{fig4} (c) compares the coefficient $A$ vs $\gamma$ for RuO$_2$ with various strongly and weakly correlated electron systems, including heavy fermion compounds and transition metals. 
This plot facilitates comparison of the KWR across different materials, highlighting the degree of electronic correlations present. 
The KWR for RuO$_2$ with RRR = 1200 is estimated to be $A$/$\gamma^2$ = 0.2$a_0$, consistent with the universal trend. These suggest that RuO$_2$ is a weakly correlated system \cite{ling2026t}.
Aside from anisotropy of a material, a molar volume is not a universal quantity. Thus, a more generalized rescaling of the original KWR plot has been introduced \cite{hussey2005,jacko2009unified}.
Jacko $et\ al.$ \cite{jacko2009unified} expressed the ratio which we call $R_\mathrm{{KWJR}}$:
\begin{equation}
R_{\mathrm{KWJR}}=\frac{A f_{di} (n)}{\gamma}= \frac{81}{4\pi  \hbar^2 k_\mathrm{B}^2 \mathrm{e}^2}. 
 \label{eq18}
 \end{equation}
\noindent Here $f_{di} (n)\equiv n D_0^2 \langle v_{0i} \rangle \xi^2$, $n$ is the density of the conduction electron, $\langle v_{0i} \rangle$ is the $k$ space average of the Fermi velocity and $\xi \approx 1$ \cite{jacko2009unified}. 
For an isotropic Fermi-liquid of dimension $d$=3 and the curent direction $i$, $\xi$=1 and $f_{3i}$ = $[3n^7/(\pi^4 \hbar^6)]^{1/3}$.
Using the conduction electron density for RuO$_2$ of $n$= 0.129$\times$ $10^{30}\ m^{-3}$, $\gamma_{\mathrm{RuO_2}}$ = 5.07 mJ mol$^{-1}$ K$^{-2}$ = 266 kg $\mathrm{K}^{-2}\mathrm{s}^{-2}\mathrm{m}$, $f_{3i}(n)$ is estimated to be 2.28$\times$10$^{135}$ kg$^2 \mathrm{m} ^{11}$s$^{-2}$. 
Figure \ref{fig4} (d) compares the $R_{\mathrm{KWJR}}$ values with other materials.
In this plot, RuO$_2$ is characterized by an exceptionally low value of $A$ among oxides and intermetallic compounds.
 \section{CONCLUSION}
RuO$_2$ shows Fermi liquid behavior with moderately enhanced effective mass and Pauli susceptibility.
Notably, the susceptibility increases in RuO$_2$ with temperature, as often observed in $d$-electron metals \cite{collings1970magnetic,collins1971some}; such behavior cannot be explained by the energy dependence of DOS with thermal activation of the quasiparticles.
The temperature dependence in RuO$_2$ is phenomenologically well fitted with $T\mathrm{ln}(T/T_0)$ up to 400 K.
This characteristic temperature-dependent susceptibility with anisotropy is attributable to lattice-expansion-induced variation of the orbital paramagnetism.
The universality plots of the Wilson ratio and the Kadowaki-Woods ratios place RuO$_2$ as a weakly correlated Fermi liquid.

\renewcommand{\thefigure}{A\arabic{figure}}  
\setcounter{figure}{0}
\begin{figure}
	\begin{center}
		\includegraphics[scale=0.55]{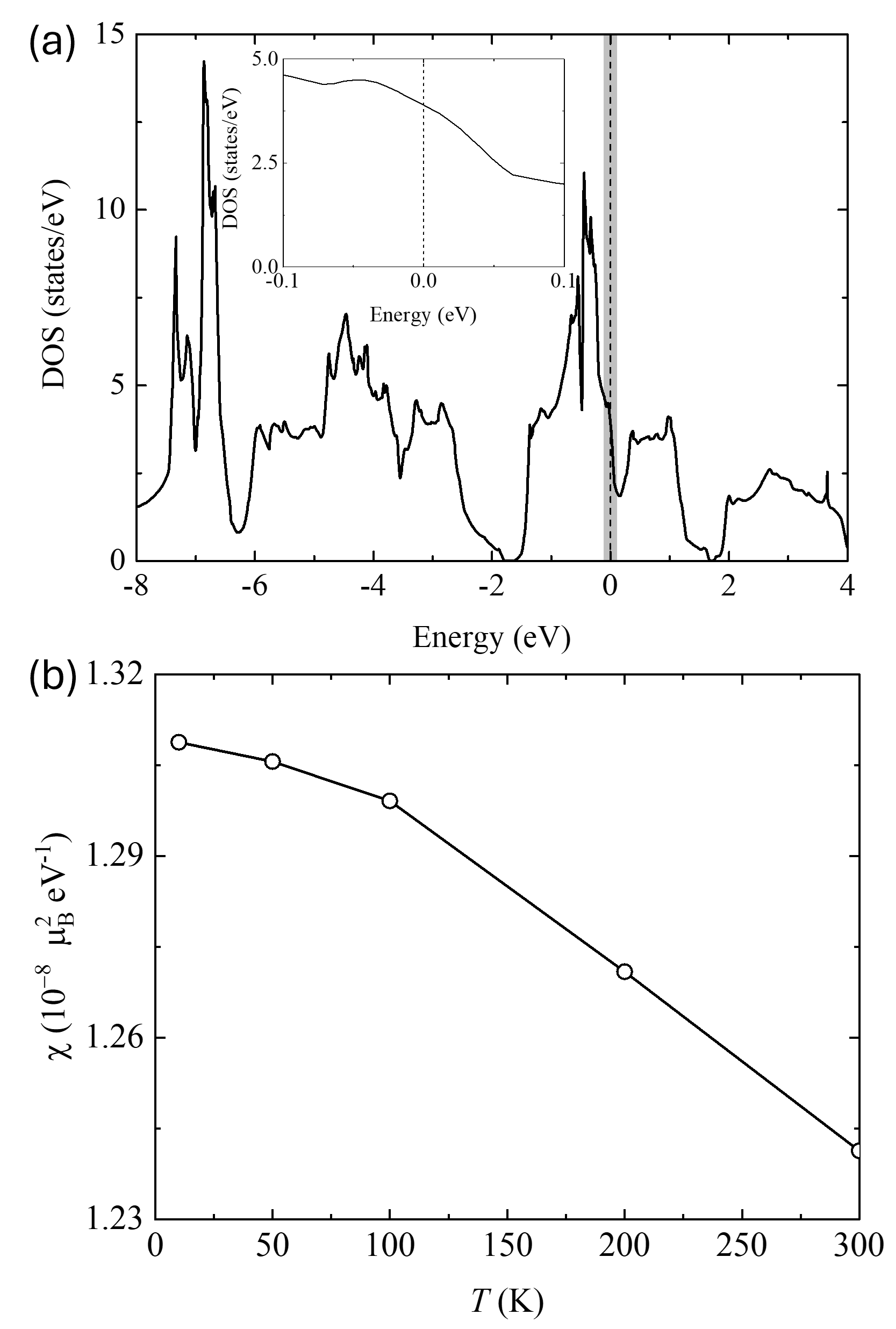}
\caption{
(a) Total density of states (DOS) of RuO$_2$ calculated within GGA ($U=0$), in the nonmagnetic state. The inset shows a magnified view of the DOS near the Fermi level.
(b) Temperature dependence of $\chi(T)$ evaluated from Eq.~(\ref{eqa01}).}
\label{A1}
    \end{center}
\end{figure}

\section{Appendix}
\subsection{Density of states (DOS) calculation}

First-principles calculations of the DOS were performed using the projector augmented-wave (PAW) method, as implemented in the Vienna ab initio simulation package (VASP) \cite{osumi2025ARPES}. 
The exchange–correlation functional was treated within the generalized gradient approximation (GGA) in the Perdew–Burke–Ernzerhof (PBE) form.
All calculations were carried out in the nonmagnetic (spin-unpolarized) state with $U=0$.
The computational setup was chosen to be consistent with our previous study \cite{osumi2025ARPES}, where the calculated band structure was directly compared with ARPES results.

Figure~\ref{A1}(a) presents the total density of states (DOS) of RuO$_2$. In the vicinity of the Fermi level, the DOS exhibits a convex shape, as highlighted in the inset.

The temperature dependence of the magnetic susceptibility was evaluated using
\begin{equation}
\chi(T) = \mu_\mathrm{B}^2 \int D(\epsilon)
\left(-\frac{\partial f(\epsilon,\mu , T)}{\partial \epsilon}\right) d\epsilon,
\label{eqa01}
\end{equation}
where $f(\epsilon,\mu,T)$ is the Fermi–Dirac distribution function.
The derivative was evaluated using the equivalent $\mathrm{sech}^2$ representation for numerical stability,
\begin{equation}
-\frac{\partial f}{\partial \epsilon}
= \frac{1}{4 k_\mathrm{B} T} \mathrm{sech}^2 \left(\frac{\epsilon}{2 k_\mathrm{B} T}\right).
\end{equation}
In the numerical integration, the energy range was restricted to $|\epsilon| < 0.5$ eV around the Fermi level.

\section*{acknowledgements}
We are grateful to S. Souma, M. Sato, T. Osumi, A. Eaton, T. Johnson, and A. V. Chubukov for useful discussions. This work was supported by the JSPS KAKENHI (JP22H01168, JP23K22439, JP25K17346) and the JST Sakura Science Exchange Program. C.S. acknowledges research support from IIT Kanpur Initiation Grant (IITK-2019-037) and research grants from Science and Engineering Research Board (SERB), Government of India (SRG2019-001104, CRG-2022-005726, EEQ-2022-000883). G.M. and H.M. acknowledge support from the Kyoto University Foundation. G.M. acknowledges support from the Toyota Riken Scholar Program. 

\bibliography{RuO2_ref}
\bibliographystyle{apsrev4-2}
\end{document}